%% file: miniJetFF.tex
\documentclass[ALICE,manyauthors]{cernphprep}

\usepackage[comma,square,numbers,sort&compress]{natbib}
\usepackage{hyperref}
\usepackage{lineno}

\usepackage{epsfig} 
\usepackage{subfigure}
\usepackage{verbatim}
\usepackage{color}
\usepackage{soul}
\usepackage{caption}  
\usepackage{array}    
\usepackage{multirow} 
\usepackage[utf8]{inputenc} 

\newcommand{\pt}{$p_{\rm T}$}
\newcommand{\ptjet}{$p_{\rm T}^{\rm ch \; jet}$}

\newcommand{\z}{$z^{\rm ch}$}
\newcommand{\negl}{\scriptsize negligible}
\newcommand{\ptnjet}{$p_{\rm T}^{\rm jet}$}

\begin{document}

\begin{titlepage}
\PHnumber{235}            
%\PHdate{\today} 
\PHyear{2018}           
\PHdate{28 August}

\title{Charged jet cross section and fragmentation \\ in
  proton-proton collisions at $\sqrt{\mathbf{s}}$~=~7~TeV}
\ShortTitle{Charged jet cross section and fragmentation at 7 TeV}   

\Collaboration{ALICE Collaboration\thanks{See Appendix~\ref{app:collab} for the list of collaboration members}}
\ShortAuthor{ALICE Collaboration} 

\begin{center}
This publication is dedicated to the memory of our colleague Oliver Busch.
\end{center}

\begin{abstract}
  We report the differential charged jet cross section and jet fragmentation distributions
measured with the ALICE detector in proton-proton collisions at a 
centre-of-mass energy $\sqrt{s}=$~7~TeV. Jets with pseudo-rapidity $\left| \eta \right| \textless {\rm 0.5}$ 
are reconstructed from charged particles using the anti-$k_{\rm T}$ jet finding algorithm with a resolution parameter  
$R$~=~0.4. The jet cross section is measured in the transverse momentum interval 5 $\leq p_{\rm T}^{\rm ch \; jet} <$ 100~GeV/$c$. Jet
fragmentation is studied measuring the scaled transverse momentum spectra of the 
charged constituents of jets in four intervals of jet transverse momentum between  5~GeV/$c$ and 30~GeV/$c$. The measurements are 
compared to calculations from the PYTHIA model as well as next-to-leading order perturbative QCD calculations
with POWHEG + PYTHIA8. The charged jet cross section is described by POWHEG for the entire measured range of \ptjet. For \ptjet $>$ 40 GeV/$c$, the PYTHIA calculations also agree with the measured charged jet cross section. PYTHIA6 simulations describe the
fragmentation distributions to 15\%. Larger discrepancies are observed for PYTHIA8. 
 \end{abstract}
\end{titlepage}
\setcounter{page}{2}

\section{Introduction}
\label{sec:introduction}

The measurement of jets in proton-proton (pp) collisions allows the study of hard scatterings and
subsequent fragmentation of partons (quarks and gluons). In this work, measurements of the charged jet cross section and jet fragmentation in pp collisions at $\sqrt{s}$ = 7~TeV are presented.  The inclusive charged jet cross section 
is measured in the transverse momentum (\pt{}) range 5~$\leq$~\ptjet{}~$<$~100~GeV/$c$. For sufficiently high \pt{}, jet
production cross sections can be calculated in perturbative
quantum chromodynamics (pQCD) supplemented with parton distribution functions (PDFs), assuming collinear
factorization. Soft processes (e.g. production of particles or prompt photons with \pt{}~$\lesssim$~2~GeV/$c$
\cite{dEnterria2014615,ALICE_neutralMesons_pp7TeV,Vogelsang_PhysRevD.48.3136}) 
cannot be described by this formalism. The measurements presented in this paper test the 
applicability of pQCD on jet production~\cite{Minijets} down to a kinematic regime of the order of a few GeV/$c$ 
and provide experimental constraints on the PDFs (see e.g.~\cite{A40_CTEQ5L}). Quantitative pQCD predictions for the charged jet cross sections
are obtained in the POWHEG~\cite{powheg1,powheg2,powheg4} scheme, in which matrix elements
are calculated at next-to-leading order (NLO) in the QCD coupling and matched
to parton shower Monte Carlo (MC) event generators to simulate parton fragmentation.

In \cite{ALICE_neutralMesons_pp7TeV}, it was found that
NLO pQCD overestimates the measured cross sections for inclusive $\pi^{0}$ and $\eta$ meson production 
at $\sqrt{s}$~=~7~TeV. Perturbative QCD calculations of hadron production rely strongly 
on parton to hadron fragmentation functions (FF) \cite{Enterria_FF}, whereas for jet
observables this dependence is much smaller. The
measured charged jet cross sections help to trace the origin of this observed discrepancy.

The production cross sections of jets in pp collisions at $\sqrt{s}$~=~7~TeV were measured previously by the
ATLAS collaboration for 100~$\leq$~\ptnjet{}~$<$~2000~GeV/$c$~\cite{ATLAS_jets5fb} and 
in the charged jet \pt{} range \\ 
4~$\leq$~\ptjet{}~$<$~100~GeV/$c$~\cite{A3_ATLASchJets} and 
by the CMS collaboration for 18~$\leq$~\ptnjet{}~$<$~1100~GeV/$c$~\cite{A5_CMSJets} and 
100~$\leq$~\ptnjet{}~$<$~2000~GeV/$c$~\cite{A5_CMSJets_1}. Jet fragmentation
in pp and Pb--Pb collisions at the LHC were reported by ATLAS~\cite{A3_ATLASchJets, A13_ATLAS_FF, ATLAS_PbPb_FF} and CMS~\cite{A14_CMS_PbPb_FF}. In \cite{ALICE_chJets7TeV}, the ALICE collaboration measured charged jet
cross sections and leading jet properties for 20$~\leq~p_{\rm T}^{\rm ch \; jet \; leading}~<~$100~GeV/$c$. 
An approximate scaling of the fragmentation distributions with the 
fractional transverse momentum \z{}~=~$p_{\rm T}^{\rm particle} / p_{\rm T}^{\rm ch \; jet}$ was observed for \z{}~$>$~0.1 and the distributions were found to be similar for the
reported $p_{\rm T}^{\rm ch \; jet \; leading}$ range. The results presented in this work repeat the previous measurements
for a jet resolution parameter of anti-$k_{\rm T}$~\cite{A25_RefAntikt} jets with R~=~0.4 with smaller uncertainties and an extended jet \pt{} coverage. 
The distributions of the fractional transverse momentum \z{} of particles in
jets with 5~$\leq$~\ptjet{}~$<$~30~GeV/$c$ presented in this work test the fragmentation scaling in a 
complementary jet \pt{} interval. Furthermore, they provide constraints on the parton shower and
hadronization models in MC event generators in a kinematic regime where strong non-perturbative effects are expected.
In commonly used event generators, soft particle production is modelled by hard parton fragmentation 
and multi-parton interactions, evoking non-perturbative colour reconnection 
\cite{Pythia_colorReconnections, Herwig_colorReconnections} at hadronization. The present results allow the perturbative contribution
to inclusive particle production to be quantified and also allow for tests of colour reconnection effects on the
fragmentation of jets with \ptjet{}~$>$~5 GeV/$c$. 

This paper is organized as follows. Section~\ref{sec:expmethod} describes
the experiment and detectors used for these measurements. 
The observables and the details of the jet reconstruction algorithms and parameters are discussed in
Sec.~\ref{sec:jetReconstruction}. 
Section~\ref{secMC} discusses the MC simulations carried out for comparisons of data to models, corrections for instrumental effects, 
and systematic uncertainty studies.  The procedures applied to correct for
instrumental effects are described 
in Sec.~\ref{sec:corrections}. The methods used to evaluate the systematic 
uncertainties of the measurements are discussed
in Sec.~\ref{sec:sysErrors}. Results are presented  and discussed in
comparison with MC Event Generator simulations in Sec.~\ref{secResults}. Section~\ref{secSummary}
summarizes the results and conclusions.

\section{Experimental setup and data sample}
\label{sec:expmethod}

The data used in this analysis were collected during the 2010 LHC pp run
with the ALICE detector~\cite{A20_AliceExpt}.  
The analysis relies primarily on the Time Projection Chamber
(TPC)~\cite{A21_RefTPC}, the Inner Tracking System (ITS)~\cite{A22_RefITS}, and the V0~\cite{A23_RefVZERO} sub-detectors. The V0
and ITS are used for event selection. The results reported in this paper are based on 177~$\times$~$\rm 10^6$ minimum bias
events corresponding to an integrated luminosity of (2.9$\pm$0.1)~nb$^{\rm -1}$~\cite{A24_AliceSigmaPaper}. 
The minimum bias trigger requires at least one hit in
either the V0 forward scintillators or in the two innermost Silicon
Pixel Detector layers (SPD) of the ITS, in coincidence with an LHC bunch crossing. 
The TPC and ITS are used for primary vertex and track reconstruction.
Only events with a primary vertex within $\pm$10~cm along the beam direction from the nominal interaction point are
analysed to minimize dependencies of the TPC acceptance on the vertex position. \par

Charged tracks are reconstructed using the combined information from the
TPC and the ITS within $\left| \eta \right| < $~0.9 over the full azimuth ($\varphi$). The track selection 
criteria are the same as described in ~\cite{ALICE_chJets7TeV} and are briefly
outlined here. 
To assure a uniform $\varphi$ distribution,
a hybrid reconstruction technique is utilized,
combining two distinct track classes: (i) tracks
containing from three to six hits in the ITS, including at 
least one hit in the SPD, and (ii) tracks containing fewer than three hits 
in the ITS, or no hit in the SPD. 
The momentum of tracks of class (i) is determined without a vertex
constraint. The vertex constraint is added for class
(ii) tracks to improve the determination of their transverse momentum.
The track momentum resolution ${\rm \delta} p_{\rm T}/p_{\rm T}$ is approximately   
4\% at \pt{}~=~40~GeV/$c$ for 95\% of all tracks. 
For tracks without a hit in the ITS (5\% of the track sample) the resolution 
is 7\% at \pt{}~=~40~GeV/$c$. Tracks from primary particles are selected 
requiring a minimum Distance of Closest Approach (DCA) to the primary vertex of 2.4 cm in the plane transverse to the beam and 3.2 cm in the beam direction.

To ensure good momentum resolution, tracks in the TPC are selected requiring a \pt{}-dependent minimum number of space points and
a maximum $\chi^2$ to ensure track fit quality. In addition, there is an upper threshold on the $\chi^2$ between the results  
of the track fit using all the space points in the ITS and TPC, and using only the TPC space points with the primary vertex position as 
an additional constraint.  

The track reconstruction efficiency for primary charged particles is
approximately 60\% at \pt{}~=~0.15~GeV/$c$, about 87\% at 1~GeV/$c$, and is nearly uniform 
up to 10~GeV/$c$ beyond which it decreases slightly. The efficiency is roughly uniform in azimuth and
within the pseudorapidity range $\left| \eta \right| <$~0.9. Further details on the track selection
procedure and tracking performance can be found in~\cite{ALICE_chJets7TeV, A15_FullJetPaper}.

\section{Jet reconstruction and observables}
\label{sec:jetReconstruction}

The anti-$k_{\rm T}$~\cite{A25_RefAntikt} algorithm from the FastJet
package~\cite{A27_RefFastjet} is used for charged jet reconstruction. Jets with a resolution parameter $R$~=~0.4 are reconstructed from charged tracks
with \pt{}~$>$~0.15~GeV/$c$ and within $\left| \eta \right| <$~0.9. The analyses reported in this work are 
restricted to jets detected within the fiducial acceptance $\left| \eta \right| <$~0.5. A boost invariant
\pt{} recombination scheme is used to determine the transverse momenta of jets 
as the sum of their charged particle transverse momenta.   

The cross section is evaluated with:
\begin{equation}
  \frac{{\rm {d^{2}}}\sigma^{\rm ch \; jet}}{{\rm d} p_{\rm
      T}\rm{d}\eta} (p_{\rm T}^{\rm ch \; jet}) =
  \frac{1}{\mathcal{L}^{\rm int}}\frac{ \rm{\Delta}{\it N}_{\rm
      jets}}{\rm {\Delta} {\it p}_{\rm T} \rm{\Delta} \eta} (p_{\rm
    T}^{\rm ch \; jet}), 
  \label{xsec-equation}
\end{equation}

where $\mathcal{L}^{\rm int}$ is the integrated luminosity and  $\rm{\Delta}{\it N}_{\rm jets}$ 
the number of jets in the selected intervals of $\rm{\Delta} {\it p}_{\rm T}$ and $\rm{\Delta} \eta$. \par

The jet fragmentation is reported based on the distribution 
\begin{equation}
  F^{z}(z^{\rm ch},p_{\rm T}^{\rm ch \; jet}) = \frac{1}{N_{\rm jets}} \frac{{\rm d}N}{{\rm d}z^{\rm ch}}, 
\end{equation}
where N is the number of charged particles. The scaled \pt{} variable \z{} 
is calculated jet by jet for each track. This observable characterizes the longitudinal jet fragmentation parallel to the jet axis.

\section{Monte Carlo simulations}
\label{secMC}

Simulations of the ALICE detector performance for particle detection and jet reconstruction are used to correct the measured distributions for instrumental effects. 
Simulated events are 
generated with the PYTHIA 6.425~\cite{Pythia} (tune Perugia-0~\cite{A31_PerugiaTunes}) MC model and 
particles are transported with GEANT3~\cite{A32_RefGeant3}. The simulated and real 
data are analysed with the same reconstruction algorithms.

In \cite{ALICE_chJets7TeV}, the detector response to simulated charged jets with \ptjet{}~$\geq$~20 GeV/$c$ was presented. In
this work, the kinematic range of jet reconstruction is extended 
to \ptjet{}~$\geq$~5 GeV/$c$. The response for low- and high-\pt{} jets was compared in the simulations. 
Consistent values were found for the \ptjet{} resolution for 5~$\leq$~\ptjet{}~$<$~20~GeV/$c$
and \ptjet{}~$\geq$~20~GeV/$c$. 

The MC models HERWIG~6.510~\cite{Herwig,A36_Herwig_1} and several PYTHIA6 tunes
are used for systematic investigations of the sensitivity of the MC correction factors to variations of the detector 
response as well as to jet fragmentation and hadronization patterns (as described in Secs. \ref{section:SysErrEffRes}
and \ref{section:SysErrUnfolding}). For comparison to data, PYTHIA6, PYTHIA8 \cite{Pythia8}, and POWHEG+PYTHIA8 simulations are used. 

PYTHIA and HERWIG are leading order (LO) event generators based on pQCD calculations of (2$\to$2) hard scattering elements. Higher order emissions are
included in the parton shower. PYTHIA and HERWIG utilize different approaches to describe the parton shower and
hadronization processes. HERWIG makes angular 
ordering a direct part of the evolution process and thereby takes coherence effects into account in the emission of soft gluons. 
PYTHIA6.4 is based on transverse-momentum-ordered showers~\cite{A37_pt_ordered_showers} in which angular ordering is imposed by an 
additional veto. In PYTHIA6 the initial-state evolution and multiple parton-parton interactions
are interleaved into one common decreasing \pt{} sequence. In PYTHIA8 the final-state evolution is also interleaved with initial state radiation and multiparton interactions. Hadronization in PYTHIA proceeds via string breaking as described by the Lund model~\cite{A38_Lund_model}, whereas
HERWIG uses cluster fragmentation. 

The PYTHIA Perugia tune variations, beginning  with the central tune Perugia-0~\cite{A31_PerugiaTunes}, 
are based on LEP, Tevatron, and SPS data. The PYTHIA6 Perugia-2011 family of tunes~\cite{A31_PerugiaTunes}
belongs to the first generation of tunes that use LHC pp data at $\sqrt{s}$~=~0.9 and 7~TeV.  For the PYTHIA8
Monash tune  \cite{Pythia8Monash} data at $\sqrt{s}$~=~8 and 13~TeV are also used. The HERWIG generator version and PYTHIA
tunes used in this work utilize the CTEQ5L parton distributions~\cite{A40_CTEQ5L}. The PYTHIA8.21 Monash tune
uses the NNPDF2.3 LO set \cite{PDF_NNPDF}.       

The POWHEG Box framework~\cite{powheg2,powheg4}, an event-by-event MC, was used for pQCD calculations of (2$\to$2) and (2$\to$3)
parton scattering at $\mathcal{O}(\alpha_{S}^3)$ in the strong coupling constant. 
The outgoing partons 
from POWHEG are passed to PYTHIA8 event-by-event where the subsequent parton shower is handled. The MC approach has the advantage that the same
selection criteria and jet finding algorithm can be used on the final state particle-level as used in the analysis of the real data; in
particular, charged particles can be selected. For the comparison with the measured differential jet cross sections,
the CTEQ6M parton distribution functions \cite{PDF_CTEQ6} are used \cite{LHAPDF5}. The dominant uncertainty in the parton-level calculation
is given by the choice of renormalization scale, $\mu_{\rm R}$, and factorization scale, $\mu_{\rm F}$. The default value was chosen
to be $\mu_{\rm R}$ = $\mu_{\rm F}$ = \pt{} of the underlying Born configuration, here a QCD 2$\to$ 2
scattering \cite{powheg4}. Independent variations by a factor of two around the central value are considered as the
systematic uncertainty. In addition, the uncertainty on the parton distribution functions has been taken into account by the variation of
the final results for the respective error sets of the parton density functions (PDFs). For the POWHEG calculations, PYTHIA8 tune Monash
was used. For test purposes, the calculations were repeated with multiparton interactions (MPI) switched off as an alternative setting.

\section{Corrections }
\label{sec:corrections}

The measured jet spectra and fragmentation distributions are corrected 
to the primary charged particle-level, as discussed in the following sections. 

\subsection{Unfolding}
\label{sec:unfolding}

Momentum-dependent imperfections in the particle detection efficiency and the finite track momentum resolution of the detector affect the jet energy scale and jet fragmentation distributions reported in 
this work. A detector response matrix is used to correct the jet spectra and fragmentation distributions for these effects. 
The instrumental response is modelled in a full simulation of the ALICE detector. Simulated events are 
generated with PYTHIA and the produced particles are transported with GEANT3. 
Jets are reconstructed both directly from the charged particle momenta produced by the MC generators ({\it particle-level}) and from the generator outputs processed through GEANT and the ALICE reconstruction software ({\it detector-level}). \par

The jet production cross sections and fragmentation distributions are corrected by
1- and 2-dimensional Bayesian unfolding \cite{A43_unfold-bayes}, respectively,
as implemented in the RooUnfold \cite{A45_RooUnfoldHtml} software.
For the unfolding of the jet cross sections, a 2-dimensional response matrix of particle-level versus detector-level charged jet \pt{}
is used. The entries of the response matrix are computed pairing particle and detector-level jets geometrically, 
according to the distance $d=\sqrt{\Delta \eta^{2} + \Delta \phi^{2}}$ between the jet axes. A bijective
match with $d<$~0.3 is required. At the smallest jet \pt{} presented in this work, \ptjet{}~=~5~GeV/$c$, the 
combined efficiency of jet reconstruction and matching detector and particle-level jets exceeds 95\%, 
and rises as a function of \ptjet{} to reach a value $>$99\% at \ptjet{}~=~20~GeV/$c$.  
The fragmentation distributions are corrected with a 4-dimensional response matrix with the axes corresponding to
particle and detector-level charged jet \pt{} and particle and detector-level \z{}, respectively.
Particle-level and detector-level jet constituents used in the calculation of \z{} are associated by matching the 
simulated TPC clusters on tracks to hits along the particle trajectories.
\par

In the Bayesian approach, the unfolding solution is regularised by the choice of the number of iterations. We observe
that the unfolded distributions typically converge to a solution after 5 steps. 
To avoid biases for the lowest and highest values of jet \pt{} reported in this paper, a wide
range  0~$<$~\ptjet{}~$<$~200~GeV/$c$ is chosen for the uncorrected distributions. 
Consistency of the unfolding procedure is ensured by folding the solution to the detector-level and comparing
it to the uncorrected distribution used as input. As an additional cross check, the analysis of charged jet cross
sections is also carried out with the RooUnfold implementation of the Singular Value
Decomposition unfolding technique ~\cite{A44_unfold-svd, A45_RooUnfoldHtml}. Consistent results are obtained with both methods. \par

The requirement of a match between the simulated detector- and particle-level jets used to compute the response matrix
introduces a kinematic bias. The effect is largest for the fragmentation distribution observable, where it is 
of the order of 5\% for small values of \z{} and 5~$\leq$~\ptjet{}~$<$~10~GeV/$c$. The bias decreases for higher values of \z{} and
\ptjet{}. For the jet cross section observable it is less than 0.5\%. We account for this effect by
applying a correction to the measured distributions prior to unfolding. The correction and the unfolding procedure are  
validated by MC closure checks, which will be discussed in Sec.~\ref{sec:sysErrors}.

\subsection{Contamination from secondary particles}
\label{sec:secondaries}

Secondary charged particles are produced by weak decays of strange particles
(e.g. $K^{\mathrm 0}_{\mathrm S}$ and $\Lambda$), decays of charged pions,  
conversions of photons from neutral pion decays and hadronic interactions in the detector material.
Although the contribution of secondaries is minimized by the track selection described in Sec.~\ref{sec:expmethod}, the measured distributions nonetheless must be corrected for a small
residual contamination.  \par

The correction for secondary particle contamination is implicitly included in the unfolding of the measured cross sections.
It is however carried out separately and explicitly prior to unfolding in the
measurements of the fragmentation function, following the procedure described in \cite{ALICE_chJets7TeV}. 
The contribution of secondaries is estimated from MC simulations,
separately for each bin in jet \ptjet{} and particle \z{}. The explicit subtraction allows for the enhancement of the low strangeness
yield in the PYTHIA Perugia-0 simulations 
to the level observed in data.  
Strange particle production in non-single-diffractive events by the CMS collaboration~\cite{A46_CMSStrangeness} and MC simulations from~\cite{Karneyeu:2013aha,A47_mcplots} are compared.  The MC predictions are scaled up to match the data.
The contamination of secondaries from strange particle decays is small, and the effect of the 
strangeness scaling on the final result is less than 1\%. 

\subsection{Underlying Event subtraction}
\label{sec:underlyingEventSub} 

The Underlying Event (UE) corresponds to all particles in an event that are not produced directly by the hard scattering of
partons. UE particles emitted in the jet cone contribute to the reconstructed jet \pt{}. To estimate and subtract the
UE activity, we use the approach discussed in \cite{ALICE_chJets7TeV}. The UE particle yield is measured event-by-event
based on circular regions transverse to the axis of the leading (highest \pt) jet. The circular regions
have the same radius as the jet resolution parameter and are 
placed at the same pseudorapidity as the leading jet but offset at an azimuthal angle ${\rm \Delta} \varphi = \pi$/2 relative to the
jet axis. For the jet cross section measurements, the UE is subtracted 
on a jet-by-jet basis prior to unfolding. The relative UE contribution to the total measured jet \pt{} is largest for the soft jets. 
The correction results in a reduction of the uncorrected jet yield by approximately 25\% for \ptjet{}~=~5~GeV/$c$ and
by about 10\% for \ptjet{}~=~20~GeV/$c$.  

The method used in \cite{ALICE_chJets7TeV} to correct the fragmentation distributions in jets
with $p_{\rm T}^{\rm ch \; jet \; leading}~\geq$~20~GeV/$c$ for the UE
applies a subtraction on the level of the constituent spectra, but does not include a simultaneous
correction to \ptjet{}. For low-\pt{} jets, this approximation may not be valid. Therefore, in this work the
fragmentation distributions are presented without correction for the UE.

\section{Systematic uncertainties}
\label{sec:sysErrors}

A summary of all systematic uncertainties for the cross section and fragmentation measurements
is given in Table~\ref{sysTable} for selected bins in \ptjet{} and \z{} to illustrate the range of systematic uncertainties. \par

\subsection{Tracking efficiency and resolution}
\label{section:SysErrEffRes}

Uncertainties associated with the momentum resolution and charged track reconstruction 
efficiency lead to systematic uncertainties in measurements of the  jet cross section
and jet fragmentation distributions. The relative systematic uncertainty on tracking efficiency is estimated to be
4\% based on variations of track selection criteria. The track momentum resolution 
has a relative systematic uncertainty of 20\%~\cite{A52_R_AA_momentumResolution}.

\par

The impact of the finite detector efficiency and momentum resolution on the unfolded jet cross sections and fragmentation distributions
is estimated by applying a parametrized detector response to PYTHIA 
events clustered with FastJet. The efficiency and resolution are varied independently, and a response matrix is computed for each variation. 
The measured distributions are unfolded, and the resulting variations are used to estimate the systematic uncertainties.
The systematic uncertainty on the jet cross sections related to tracking efficiency
increases smoothly with increasing \ptjet{}. For the fragmentation distributions, the uncertainty
is largest at \z{}~=~1 and has a minimum at \z{}~$\approx$~0.35. The systematic uncertainty on the measured cross sections and fragmentation
distributions from finite momentum resolution is comparatively small, and largest for high \ptjet{} and \z{}. 

\subsection{Unfolding}
\label{section:SysErrUnfolding}

The data correction methods used in this work are largely based on tune Perugia-0 of the PYTHIA event generator. The 
particular structure of jets simulated by PYTHIA might however affect the simulated detector response and influence the correlation
between particle and detector-level quantities used to compute the response matrices. Furthermore, the RooUnfold Bayesian unfolding algorithm is based on
a prior solution which is initially obtained from the MC and updated in subsequent iterations. The choice of a
particular initial prior might have an impact on the unfolded solution. Such event generator dependencies are examined by comparing unfolded solutions obtained with response matrices from the PYTHIA
tunes Perugia-0 and Perugia-2011 with those obtained with the HERWIG generator. This is accomplished with a parametrized 
detector response and the anti-$k_{\rm T}$ jet finder. The resulting systematic uncertainties on the jet cross sections
are largest for the lowest \ptjet{}. For the fragmentation distributions, the strongest event generator dependence is observed for
the lowest jet \pt{}, in the interval 5~$\leq$~\ptjet{}~$<$~10~GeV/$c$, where the uncertainty is largest for intermediate values of \z{}~$\approx$~0.4 and for \z{}~=~1.
The distributions for \ptjet{}~$\geq$~10~GeV/$c$ show a monotonic increase of the systematic uncertainty with \z{}.  \par

The unfolding approach is validated by closure tests on PYTHIA simulations. To detect potential biases, the simulated detector-level distribution
is unfolded and the solution is compared to the particle-level truth. For the unfolded jet cross section, no significant difference is observed. For the
fragmentation distributions, a small systematic bias can be detected. We assign a constant uncertainty of 1\% to account for this non-closure.

\subsection{Correction for secondary charged particles}

The systematic uncertainty associated to the correction for the contribution from secondary charged particles to
the jet cross sections and fragmentation distributions is estimated by varying track selection criteria. We 
change the contribution of secondary charged particles by varying the track selection criteria~\cite{ALICE_chJets7TeV} and
correct the measured distributions accordingly. Residual variations of the corrected distributions are used
to estimate the systematic uncertainties. The resulting uncertainties on the fragmentation distributions are largest at
small values of \z{}. 
The uncertainty on the measured jet cross section is evaluated as a \ptjet{} scale uncertainty of 0.5\%. 
 
\subsection{Underlying Event subtraction}

The jet cross sections are corrected for the contribution from the UE. In \cite{ALICE_chJets7TeV}, the uncertainty on the measurement of the
UE \pt{} density was estimated to be 5\%. The corresponding uncertainty of the jet cross section is evaluated as a jet \pt{} scale uncertainty 
resulting in a systematic uncertainty which is 2\% for \ptjet{}~=~5~GeV/$c$ and decreases for higher \ptjet{}.

\par

\newcommand{\specialcell}[2][l]{\begin{tabular}[#1]{@{}l@{}}#2\end{tabular}}  

\renewcommand{\arraystretch}{1.5}
\begin{table}[b]

\small 

\centering  

\begin{tabular}{ 
   m{2.0cm}             |  
   r                       
   m{1.45cm}<{\centering}  
   m{1.25cm}<{\centering}  
   m{1.25cm}<{\centering}  
   m{1.25cm}<{\centering}   
   m{1.25cm}<{\centering}   
   m{0.9cm}<{\centering}   
   m{0.9cm}<{\centering}  
   m{1.0cm}<{\centering} } 
 
 \hline \hline 
 
\mbox{Distribution }  & Bin \rule{0.5cm}{0cm}& Track eff. (\%) & Track \pt{} res. (\%)& Event Generator (\%) & MC Closure (\%) & Sec. corr. (\%) & UE (\%) &  Norm. (\%) & Total (\%) \\ 
\hline 

\multirow {3}{*}{   $ \displaystyle \frac{ { \rm {d^{2}}}\sigma^{\rm ch \; jet} } { {\rm d} p_{\rm T}^{\rm ch \; jet}\rm{d}\eta }$  }
&  5-6~GeV/$c$   &  7.0 & 0.1 & 2.1 & - & 2.0 & 1.7 & 3.5 &  8.6  \\ \cline{2-10}
&  20-24~GeV/$c$ & 10.2 & 0.5 & 1.0 & - & 2.2 & 0.5 & 3.5 & 11.1  \\ \cline{2-10} 
& 86-100~GeV/$c$ & 11.7 & 2.0 & 1.0 & - & 2.6 & 1.5 & 3.5 & 12.7  \\ \cline{1-10}

\multirow {3}{*}{ \specialcell[c]{{ $ \displaystyle \frac{1}{N_{\rm jets}} \frac{{\rm d}N}{{\rm d}z^{\rm ch}}$} \\ {\tiny 5$\leq p_{T}^{\rm ch \; jet}<$ 10 GeV/$c$}
  }
}  
       & 0   - 0.1   &  4.1 & \negl & 1.4 & 1.0 & 3.2 & - & - &  5.5 \\ \cline{2-10}
       & 0.35 - 0.4  &  0.1 & 0.2   & 2.9 & 1.0 & 0.6 & - & - &  3.2 \\ \cline{2-10}
       & 0.95 - 1.0  & 10.4 & 0.6   & 4.7 & 1.0 & 0.2 & - & - & 11.4 \\ \cline{1-10}

\multirow {3}{*}{ \specialcell[c]{$ \displaystyle \frac{1}{N_{\rm jets}} \frac{{\rm d}N}{{\rm d}z^{\rm ch}}$ \\ {\tiny 15$\leq p_{T}^{\rm ch \; jet}<$ 20 GeV/$c$}
} } 
       & 0   - 0.1   &  4.0 & \negl & 0.6 & 1.0 & 2.6 & - & - &  4.9 \\ \cline{2-10}
       & 0.35 - 0.4  &  7.9 & 0.7   & 0.9 & 1.0 & 0.4 & - & - &  1.7 \\ \cline{2-10}
       & 0.95 - 1.0  &  9.0 & 1.9   & 2.5 & 1.0 & 0.5 & - & - &  9.6 \\ \cline{1-10}

\hline 
\hline
\end{tabular}
\captionsetup{width = 1.0\textwidth}
\caption{\normalsize Summary of systematic uncertainties of the cross section and fragmentation distributions for selected bins in \ptjet{} and \z{}. The contributions from tracking efficiency and track \pt{} resolution, the event generator dependence
  of the unfolding correction, MC closure, secondaries correction, UE subtraction and cross section
  normalization as well as the total uncertainty are shown.}

\normalsize 

\label{sysTable}
\end{table}

\clearpage

\section{Results}\label{secResults}

\begin{figure*}[b]
 \begin{center}
   \includegraphics[width=0.5\textwidth]{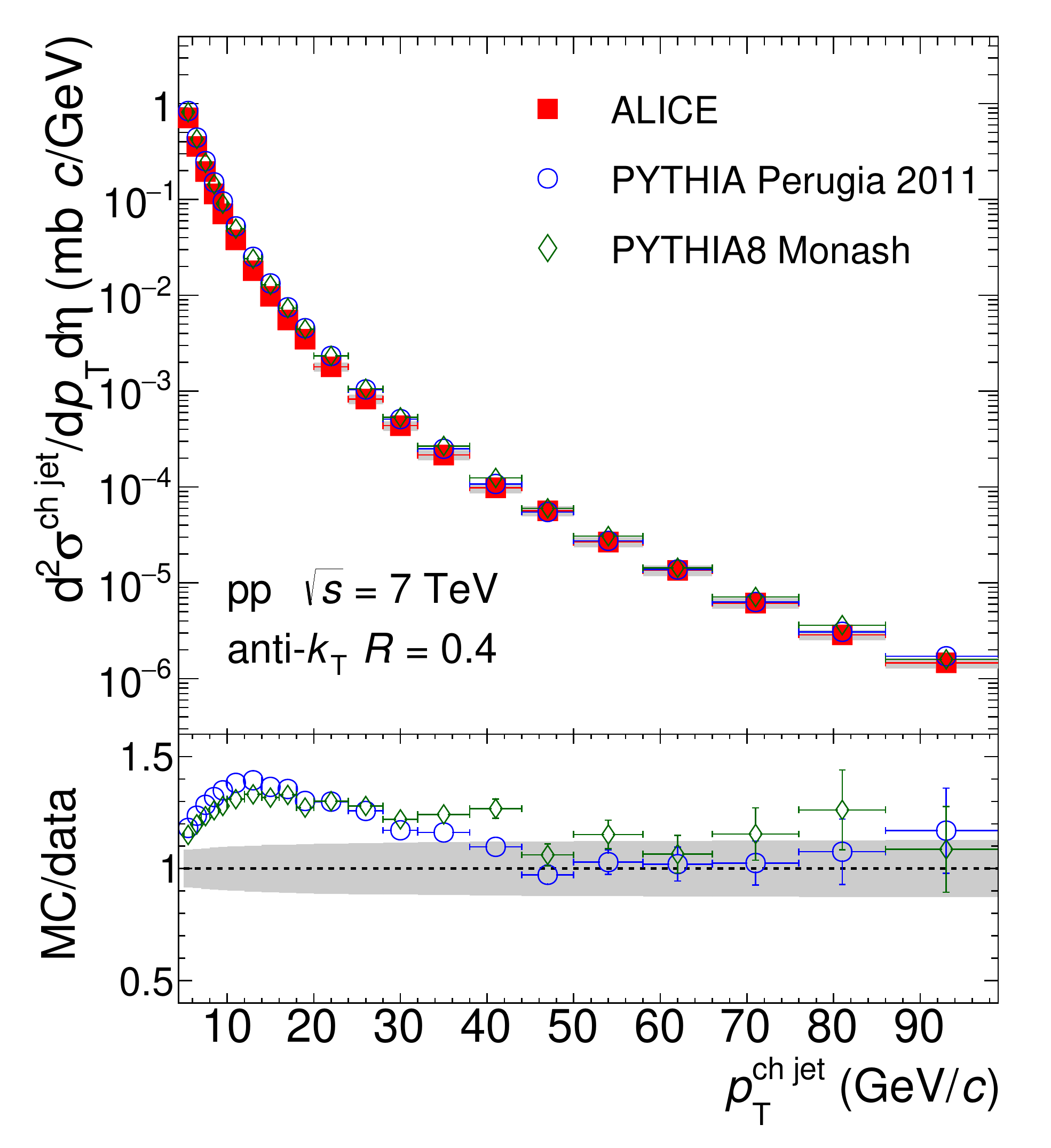} 
   \caption{Top panel: inclusive charged jet cross section in pp collisions at $\rm{\sqrt{s}=7}$~TeV using 
     the anti-$k_{\rm T}$ algorithm with $R$~=~0.4 compared to calculations from PYTHIA6 Perugia-2011
     and PYTHIA8 tune Monash. Bottom panel: ratios of MC distributions to data. The shaded band shows the 
     systematic uncertainty on the data drawn at unity, error bars represent the statistical uncertainties.
     Most uncertainties are smaller than the marker size. }
   \label{Fig:xsec_compMC}
 \end{center}
\end{figure*}

Figure~\ref{Fig:xsec_compMC} presents the inclusive charged jet cross section measured in 
pp collisions at $\rm{\sqrt{s}~=}$~7~TeV using the anti-$k_{\rm T}$ jet finder. The cross section is reported 
for a resolution parameter $R$~=~0.4 in the pseudo-rapidity interval $\left| \eta \right| < $~0.5.
Statistical uncertainties are displayed as vertical error bars. The total systematic uncertainties are obtained as a quadratic sum of
the individual contributions described in Sec.~\ref{sec:sysErrors}, and are shown as shaded boxes around the data points. 
The results presented in this work extend the jet \ptjet{} coverage of previous measurements of the charged jet cross section 
by the ALICE collaboration \cite{ALICE_chJets7TeV}, with reduced systematic uncertainties, and are consistent in the common \ptjet{} range.
The previous results are superseded by this work. 

The measured charged jet cross sections are compared to calculations from the PYTHIA MC model. The ratios of the
MC distributions to measured data are shown in the bottom panel. The 
systematic uncertainty on the data is indicated by a shaded band drawn at unity. The models qualitatively describe the
measured cross sections, but fail to reproduce the spectral shape over the entire range of measured jet \ptjet{}.
In the high jet transverse momentum range, \ptjet{}~$>$~40~GeV/$c$ 
both PYTHIA6 tune Perugia-2011 and PYTHIA8 tune Monash describe  
the data well, whereas at intermediate \ptjet{} the jet cross section is systematically
overestimated. The discrepancy is about 30-40\% for \ptjet{}~$\approx$~10-15~GeV/$c$.   

\begin{figure*}[htb]
 \begin{center}
   \includegraphics[width=0.5\textwidth]{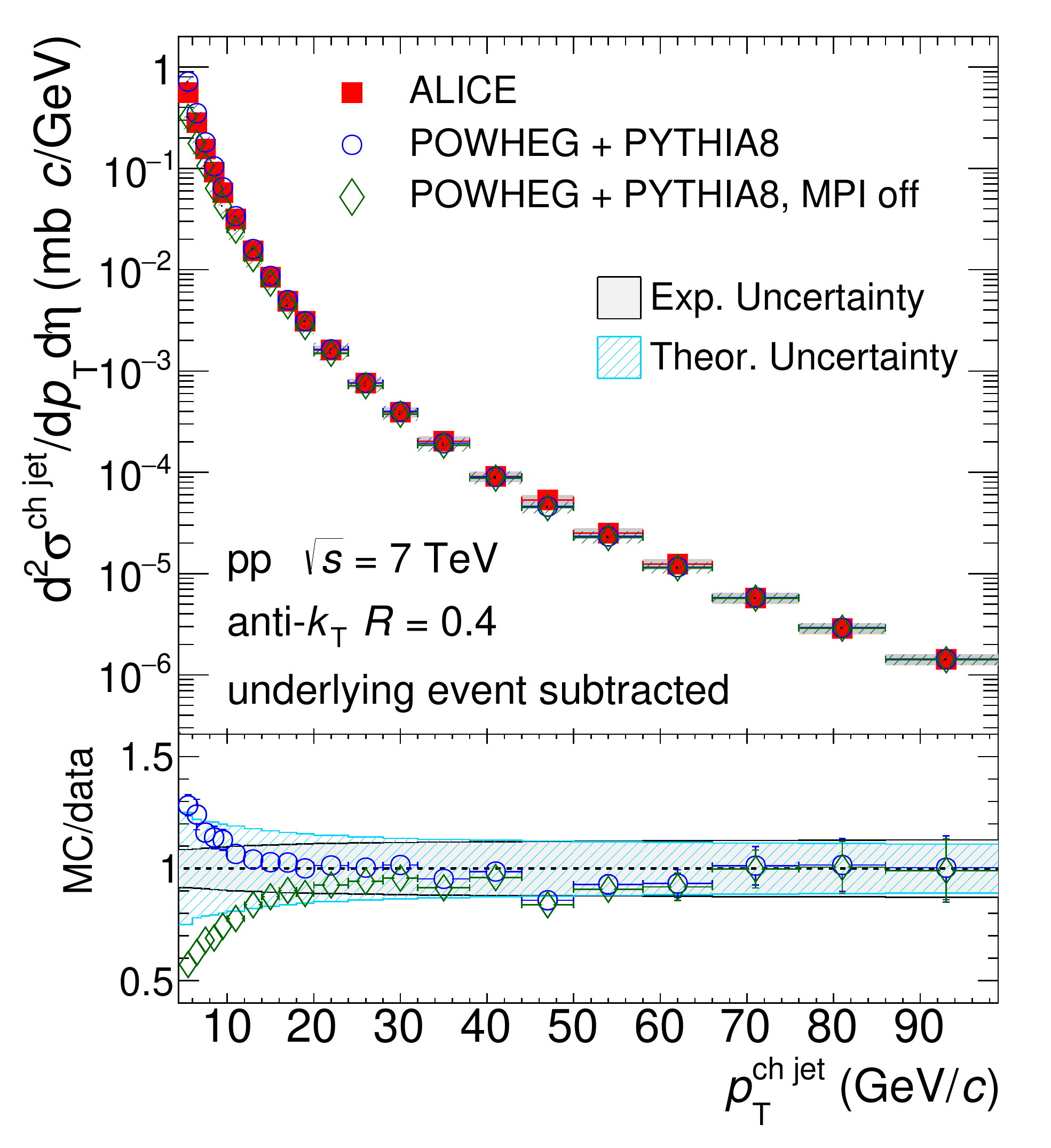} 
   \caption{Top panel: inclusive charged jet cross section compared to POWHEG + PYTHIA8 NLO pQCD calculations with and without
     MPI. In data and calculations, the Underlying
     Event contribution is subtracted. Bottom panel: ratio of POWHEG calculations to data. The shaded bands indicate systematic uncertainties
     on data and theory predictions.} 
   \label{Fig:xsec_compPOWHEG}
 \end{center}
\end{figure*}

In Fig.~\ref{Fig:xsec_compPOWHEG}, the measured cross sections are compared to NLO pQCD calculations with
the POWHEG Box framework, in which the outgoing partons are passed to PYTHIA8 where the subsequent parton shower and hadronization
are handled. The UE contribution is subtracted using the method described
in Sec.~\ref{sec:underlyingEventSub} in both data and theory calculations. Systematic uncertainties on data and theory predictions
are indicated by shaded bands. The theory systematic uncertainties are related to the choice of scale and PDF as well as the
UE subtraction. They are largest at the lowest \ptjet{} and vary between 25\% and 11\%. In the jet transverse
momentum range \ptjet{}~$>$~7~GeV/$c$, POWHEG + PYTHIA8 (open circles) gives a good description of the data. The spectral shape
is reproduced well for \ptjet{}~$>$~20~GeV/$c$. At lower transverse momenta, 5~$\leq$~\ptjet{}~$<$~20 GeV/$c$, the calculations overestimate the measured cross section, but the difference is within the combined experimental and theoretical
uncertainties. 
To study the contribution of soft processes generated in PYTHIA8, the POWHEG + PYTHIA8 calculations were repeated with 
alternative settings, switching off MPI from PYTHIA8. The calculated jet cross sections 
without MPI (open diamonds) are smaller than the result with default settings in the range \ptjet{}~$<$~20 GeV/$c$, and
the measured jet cross section is significantly underpredicted for \ptjet{}~$<$~10~GeV/$c$. The agreement with the data
is worse than in the case with MPI.  As a further test, we compared the UE activity, measured by the 
particle \pt{} density in perpendicular cones, in data and simulations for default and alternative settings. The POWHEG + PYTHIA8 simulations
with default settings reproduce the measured UE reasonably well (compare also \cite{Pythia8Monash}), whereas 
simulations without MPI show a strongly reduced UE \pt{} density and fail to describe the data.
These results indicate a sizable contribution from non-perturbative processes to jet production at low \ptjet{}. Comparing the
two settings in the simulations, MPI contribute $\sim$ 50\% to the cross section for 5~$\leq$~\ptjet{}~$<$~10~GeV/$c$ and $\sim$ 20\% for \ptjet{}~$\geq$~10~GeV/$c$.
In this estimate, a possible additional contribution from initial state radiation is not taken into account. In a study of low transverse energy clusters 
in p$\mathrm{\bar{p}}$ collisions at $\rm{\sqrt{s}=900}$~GeV \cite{UA1_minijets}, the contribution from soft processes to
jets with $E_{\rm T}^{\rm raw}$~$>$~5~GeV was evaluated to be 18\%, similar in magnitude but lower than our estimate. This 
difference may be attributed to experimental differences in the definition of the jet energy scale and in the theoretical models, but
may also reflect the $\sqrt{s}$ evolution of the probability for MPI, represented by the rise of the UE density observed 
with collision energy \cite{A51_ALICE_UE}. 

Next-to-leading order pQCD calculations overestimate the cross sections for
inclusive $\pi^{0}$ and $\eta$ meson production at midrapidity measured in pp collisions at $\sqrt{s}$~=~7~TeV
in the $\pi^{0}$ ($\eta$) transverse momentum range 0.3~$<$~\pt{}~$<$~25~GeV/$c$ (0.4~$<$~\pt{}~$<$~15 GeV/$c$) by up to a factor of three~\cite{ALICE_neutralMesons_pp7TeV}.
The jet cross section presented in this work covers a \pt{} range consistent with \cite{ALICE_neutralMesons_pp7TeV}, and a consistent
PDF set was used for the POWHEG calculations. Since the jet cross section observable depends only weakly on the details of parton fragmentation,
the good agreement between data and NLO pQCD calculations for jet cross sections suggests the uncertainty in the parton to hadron fragmentation functions to be the cause for the discrepancy observed for neutral mesons.

\begin{figure*}[ht]
\rotatebox{0}{\resizebox{\textwidth}{!}{
\includegraphics[scale=0.45]{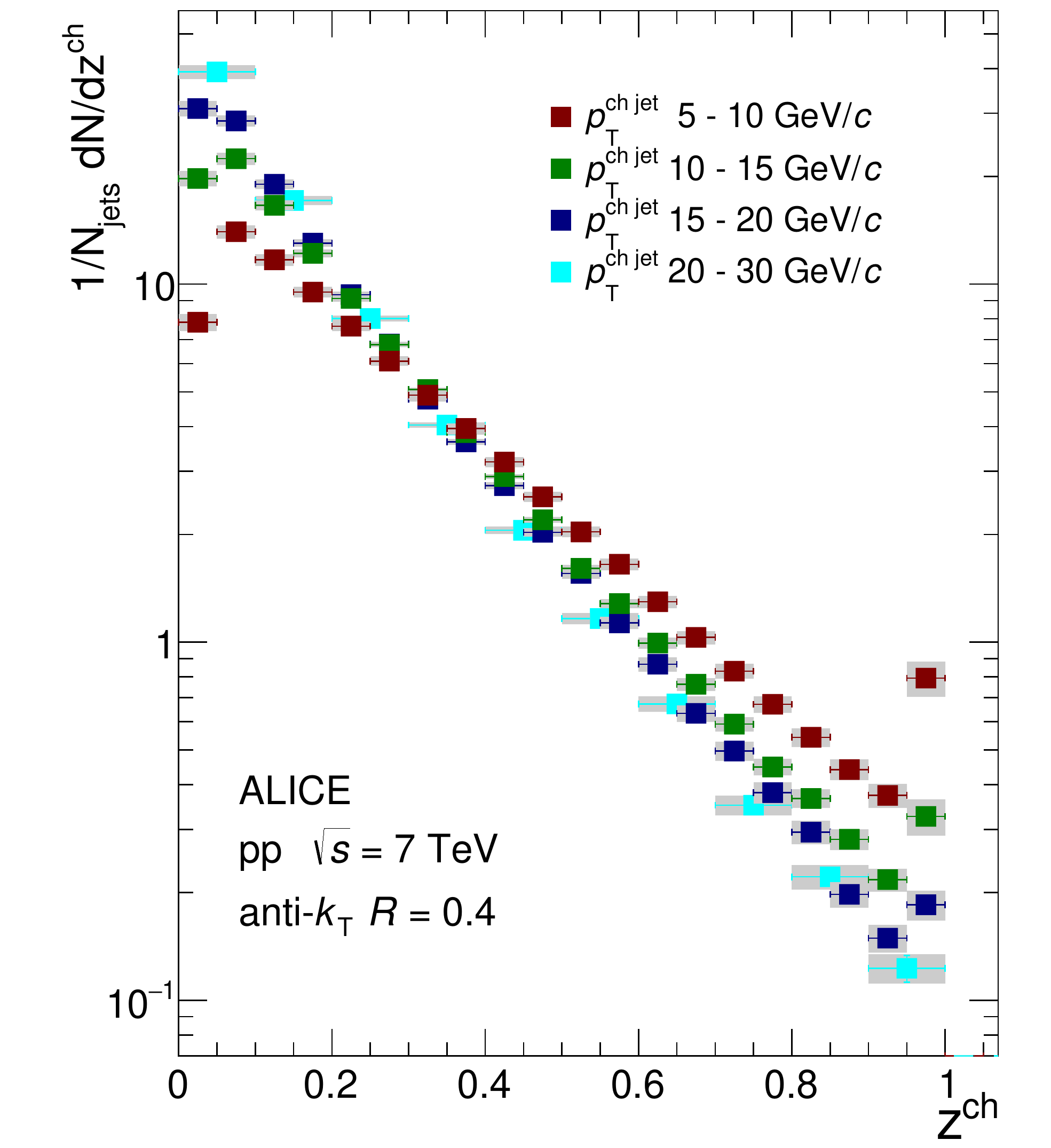}
\includegraphics[scale=0.45]{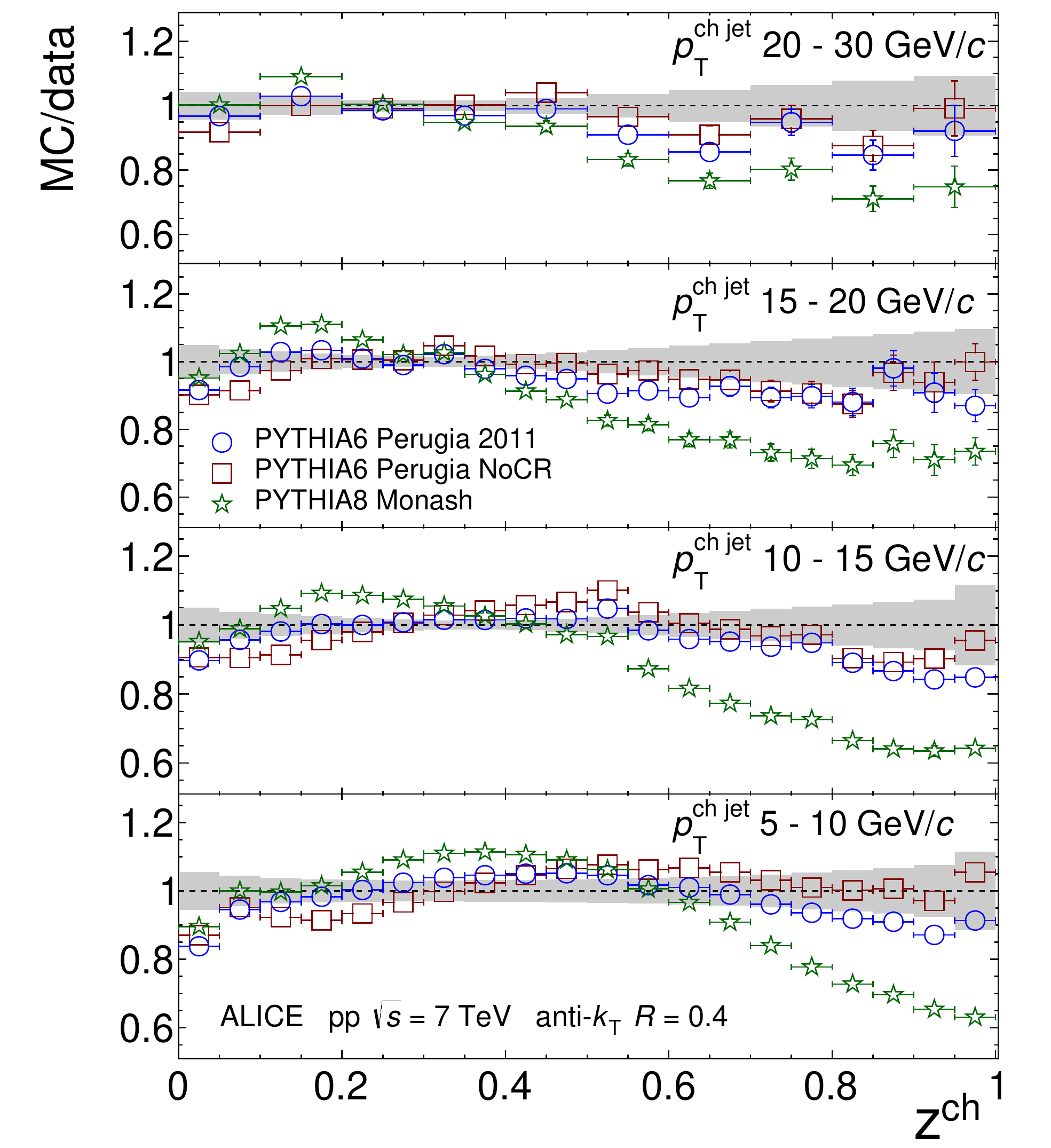}
}}
\caption{Left panel: Charged particle scaled \pt{} spectra $F^{z}(z^{\rm ch},p_{\rm T}^{\rm ch \; jet})$ for different
  bins in jet transverse momentum. Right panel: Ratio of MC distributions to data. The shaded band shows the 
  systematic uncertainty on the data drawn at unity.  Error bars represent the statistical uncertainties.}
  \label{fig:FFz}
\end{figure*}

The left panel of Fig.~\ref{fig:FFz} presents the measured 
scaled \pt{} spectra, $F^{z}$, of charged particles in charged jets reconstructed 
with a resolution parameter $R$~=~0.4. The $F^{z}$ distributions are shown for four bins in jet \pt:
5~$\leq$~\ptjet{}~$<$~10~GeV/$c$, 10~$\leq$~\ptjet{}~$<$ 15~GeV/$c$, 15~$\leq$~\ptjet{}~$<$~20~GeV/$c$ and 20~$\leq$~\ptjet{}~$<$~30~GeV/$c$. The
spectra span two to three orders of magnitude. 
At the lowest \z{}, for jets with \ptjet{}~$<$~10~GeV/$c$ the yield increases to a distinct maximum at \z{}~$\approx$~0.05. This non-monotonic behaviour corresponds
to the hump-backed plateau at high values of the variable $\xi = \mathrm{log} (1/z)$ \cite{CDF_FF, ALICE_chJets7TeV}, which reflects the suppression
of low momentum particle production
by QCD coherence~\cite{A29_QCD_coherence,A29_QCD_coherence_1}. For jets with 10 $\leq$~\ptjet{}~$<$~15~GeV/$c$ the maximum is less pronounced, and for jets
with 15~$\leq$~\ptjet{}~$<$~20~GeV/$c$ the yield is roughly constant for z~$<$~0.1. This reflects a shift of the maximum towards lower \z{} (corresponding
to higher $\xi$) with increasing \ptjet{}. A similar \ptjet{} dependence was observed in \cite{ALICE_chJets7TeV}. For the highest \z{} bin,
the $F^{z}$ distributions for jets with \ptjet{}~$<$~20~GeV/$c$ show a discontinuous increase, which is strongest 
for the lowest \ptjet{} bin. It corresponds to jets with only a single charged constituent,
for which \z{}~=~1 by construction. The effect is also observed in the simulations.

An increase of the
integral of the distributions with \ptjet{} is observed, reflecting the rise of particle multiplicity with increasing \ptjet{}. 

In \cite{ALICE_chJets7TeV} it was found 
that the $F^{z}$ distributions measured for leading jets in the range $p_{\rm T}^{\rm ch \; jet \; leading}~\geq~$20~GeV/$c$ are
constistent within uncertainties
for \z~$ > $~0.1, indicating a scaling of charged jet fragmentation with charged jet transverse momentum.
For the inclusive jet fragmentation distributions in the jet \pt{} range reported in this work, no such scaling is observed.
The shape of the spectra become progressively flatter with decreasing \ptjet{}. However, comparing $F^{z}$  for
the jet \pt{} bin 15-20~GeV/$c$ and the lowest jet \pt{} bin, 5-10~GeV/$c$, to $F^{z}$ for the intermediate
jet \pt{} bin, 10-15~GeV/$c$, we observe that the distributions for the two higher bins are more similar to each other than
the two lower \pt{} bins. This may indicate an onset of the scaling behaviour. 
We note that the distribution for inclusive jets with 15~$\leq p_{\rm T}^{\rm ch \; jet} <$~20~GeV/$c$ and
20~$\leq p_{\rm T}^{\rm ch \; jet} <$~30~GeV/$c$ exhibit small but significant differences. This indicates that
the \z{} scaling reported in \cite{ALICE_chJets7TeV} is only fully developed for 
\ptjet{}~$\geq$~20~GeV/$c$. 

The measured fragmentation distributions are compared to calculations
obtained from the PYTHIA model, and the ratios of the MC distributions to data
are presented in the right panel of Fig.~\ref{fig:FFz}. The observed trends for the individual
tunes are similar for all charged jet \pt{}. The PYTHIA6 tune Perugia-2011 reproduces the  
fragmentation distributions reasonably well, although there are discrepancies of up to 10-15\% in some kinematic regions. For \ptjet{}$~\geq~$10~GeV/$c$, the model tends to underpredict the measured yield at high \z{}.  
The PYTHIA8 calculations with the Monash tune exhibit a softer spectrum than the data,
overpredicting the fragment yield at intermediate \z{}~$\approx$~0.15~--~0.4 and 
underestimating the rates at high \z{}, the discrepancy reaching $\sim$35\% at $z=1$ for the lowest jet \pt{} bin. 
The difference between calculations and data at intermediate \z{} is most pronounced at a
value of constituent \pt{}~$\approx$~2~GeV/$c$ for all four jet \pt{} intervals. To investigate the observed 
differences between data and calculations at higher jet \pt{}, we also compared the leading charged jet $F^{z}$ distributions in the range
 20~$\leq~p_{\rm T}^{\rm ch \; jet \; leading}~<$~80~GeV/$c$ from \cite{ALICE_chJets7TeV}
to PYTHIA8 simulations. We observe that for $p_{\rm T}^{\rm ch \; jet \; leading}~\geq$~40 GeV/$c$ the distributions at intermediate \z{} are well described, whereas the yield at
high \z{} is also underestimated for high $p_{\rm T}^{\rm ch \; jet \; leading}$.  

The data are also compared to the PYTHIA6 Perugia NoCR tune~\cite{A31_PerugiaTunes}. This tune is an attempt to describe the 
data sets used for the Perugia tunes without invoking colour reconnections (CR) \cite{Pythia_colorReconnections}
between fragmenting partons to model non-perturbative colour string interactions.  It does not
reproduce the data used to constrain the PYTHIA parameter space well. However,
for the $F^{z}$ distributions reported in this paper, the calculations agree with the data to within about 10-15\%. 
In \cite{Ortiz_colorReconnections} it was shown that in the PYTHIA8 model, the effect of CR is
strong in events with MPI and increases 
with MPI activity. Hence, the weak effect of colour reconnections on the low-\ptjet{} fragmentation distributions in PYTHIA may indicate 
that these jets are dominantly produced in hard scattering events and from MPI with a few hard outgoing partons, rather than being formed as 
hadron clusters from the fragmentation of many soft partons combined by the jet finding algorithm. 

\section{Summary} \label{secSummary}

The inclusive charged jet cross section and jet fragmentation distributions at midrapidity in
pp collisions at $\sqrt{s}$~=~7~TeV were measured. The cross section for a resolution parameter $R$~=~0.4
was reported in the \ptjet{} interval from 5 to 100~GeV/$c$. We studied charged particle fragmentation in charged jets with 5~$\leq$~\ptjet{}~$<$~30~GeV/$c$, extending the range in~\cite{ALICE_chJets7TeV}.
The integral of the fragmentation distributions increases with jet \pt{}, showing an increase of
particle multiplicity in jets. The shape of the distributions become progressively flatter for lower jet \pt{}.

The measurements were compared to PYTHIA calculations. The cross sections are well described by PYTHIA6 and PYTHIA8 for \ptjet{}~$>$~40~GeV/$c$. 
At lower \ptjet{} the PYTHIA tunes studied here fail to describe the shape of the jet spectra and the cross section is systematically overestimated.
PYTHIA6 tune Perugia-2011 gives a reasonable description of the fragmentation distributions, whereas the PYTHIA8 tune Monash exhibits a
softer spectrum than the data, with significant deviations particularly at high \z.

The jet cross sections are well described by POWHEG NLO pQCD + PYTHIA8 calculations for the entire measured range of \ptjet{}. The simulations 
indicate a sizable contribution of Multi-Parton-Interactions to the jet cross section for low \ptjet{}. We found that 
PYTHIA6 tune NoCR reproduces the measured fragmentation distributions reasonably well in the entire jet \pt{} range covered by our measurements, 
possibly indicating that the contribution of events with multiple soft colour connected partons to jet production is small in the kinematic
regime of our measurement, \ptjet{}~$>$~5~GeV/$c$.   

The good agreement between the NLO calculations and the measured jet cross section indicates that the previously observed discrepancies
between data and NLO calculations of neutral meson production may be due to the fragmentation functions used in these calculations.

%%%%% acknowledgements
\newenvironment{acknowledgement}{\relax}{\relax}
\begin{acknowledgement}
\section*{Acknowledgements}
\input{fa_2018-09-03.tex}
\end{acknowledgement}

\bibliographystyle{utphys}   
\bibliography{biblio}

%%%%%%%%% appendix with author list
\newpage
\appendix
\newpage
\section{The ALICE Collaboration}
\label{app:collab}
\input{2018-06-23-Alice_Authorlist_2018-Jun-23.tex}
\end{document}

%% file: fa_2018-09-03.tex
% Version: 2018-09-03

The ALICE Collaboration would like to thank all its engineers and technicians for their invaluable contributions to the construction of the experiment and the CERN accelerator teams for the outstanding performance of the LHC complex.
The ALICE Collaboration gratefully acknowledges the resources and support provided by all Grid centres and the Worldwide LHC Computing Grid (WLCG) collaboration.
The ALICE Collaboration acknowledges the following funding agencies for their support in building and running the ALICE detector:
A. I. Alikhanyan National Science Laboratory (Yerevan Physics Institute) Foundation (ANSL), State Committee of Science and World Federation of Scientists (WFS), Armenia;
Austrian Academy of Sciences and Nationalstiftung f\"{u}r Forschung, Technologie und Entwicklung, Austria;
Ministry of Communications and High Technologies, National Nuclear Research Center, Azerbaijan;
Conselho Nacional de Desenvolvimento Cient\'{\i}fico e Tecnol\'{o}gico (CNPq), Universidade Federal do Rio Grande do Sul (UFRGS), Financiadora de Estudos e Projetos (Finep) and Funda\c{c}\~{a}o de Amparo \`{a} Pesquisa do Estado de S\~{a}o Paulo (FAPESP), Brazil;
Ministry of Science \& Technology of China (MSTC), National Natural Science Foundation of China (NSFC) and Ministry of Education of China (MOEC) , China;
Ministry of Science and Education, Croatia;
Centro de Aplicaciones Tecnol\'{o}gicas y Desarrollo Nuclear (CEADEN), Cubaenerg\'{\i}a, Cuba;
Ministry of Education, Youth and Sports of the Czech Republic, Czech Republic;
The Danish Council for Independent Research | Natural Sciences, the Carlsberg Foundation and Danish National Research Foundation (DNRF), Denmark;
Helsinki Institute of Physics (HIP), Finland;
Commissariat \`{a} l'Energie Atomique (CEA) and Institut National de Physique Nucl\'{e}aire et de Physique des Particules (IN2P3) and Centre National de la Recherche Scientifique (CNRS), France;
Bundesministerium f\"{u}r Bildung, Wissenschaft, Forschung und Technologie (BMBF) and GSI Helmholtzzentrum f\"{u}r Schwerionenforschung GmbH, Germany;
General Secretariat for Research and Technology, Ministry of Education, Research and Religions, Greece;
National Research, Development and Innovation Office, Hungary;
Department of Atomic Energy Government of India (DAE), Department of Science and Technology, Government of India (DST), University Grants Commission, Government of India (UGC) and Council of Scientific and Industrial Research (CSIR), India;
Indonesian Institute of Science, Indonesia;
Centro Fermi - Museo Storico della Fisica e Centro Studi e Ricerche Enrico Fermi and Istituto Nazionale di Fisica Nucleare (INFN), Italy;
Institute for Innovative Science and Technology , Nagasaki Institute of Applied Science (IIST), Japan Society for the Promotion of Science (JSPS) KAKENHI and Japanese Ministry of Education, Culture, Sports, Science and Technology (MEXT), Japan;
Consejo Nacional de Ciencia (CONACYT) y Tecnolog\'{i}a, through Fondo de Cooperaci\'{o}n Internacional en Ciencia y Tecnolog\'{i}a (FONCICYT) and Direcci\'{o}n General de Asuntos del Personal Academico (DGAPA), Mexico;
Nederlandse Organisatie voor Wetenschappelijk Onderzoek (NWO), Netherlands;
The Research Council of Norway, Norway;
Commission on Science and Technology for Sustainable Development in the South (COMSATS), Pakistan;
Pontificia Universidad Cat\'{o}lica del Per\'{u}, Peru;
Ministry of Science and Higher Education and National Science Centre, Poland;
Korea Institute of Science and Technology Information and National Research Foundation of Korea (NRF), Republic of Korea;
Ministry of Education and Scientific Research, Institute of Atomic Physics and Romanian National Agency for Science, Technology and Innovation, Romania;
Joint Institute for Nuclear Research (JINR), Ministry of Education and Science of the Russian Federation and National Research Centre Kurchatov Institute, Russia;
Ministry of Education, Science, Research and Sport of the Slovak Republic, Slovakia;
National Research Foundation of South Africa, South Africa;
Swedish Research Council (VR) and Knut \& Alice Wallenberg Foundation (KAW), Sweden;
European Organization for Nuclear Research, Switzerland;
National Science and Technology Development Agency (NSDTA), Suranaree University of Technology (SUT) and Office of the Higher Education Commission under NRU project of Thailand, Thailand;
Turkish Atomic Energy Agency (TAEK), Turkey;
National Academy of  Sciences of Ukraine, Ukraine;
Science and Technology Facilities Council (STFC), United Kingdom;
National Science Foundation of the United States of America (NSF) and United States Department of Energy, Office of Nuclear Physics (DOE NP), United States of America.

%% file: 2018-06-23-Alice_Authorlist_2018-Jun-23.tex
% Collaboration: CERN-LHC-ALICE
% Generated manually. System still not generating authorlist properly at CR2

% How to use:
%%%%%%%%% appendix with author list
%\appendix
%\section{The ALICE Collaboration}
%\label{app:collab}
%\input{Alice_Authorslist_XXXX-Axx-XX.tex}
\begingroup
\small
\begin{flushleft}
S.~Acharya$^{\rm 139}$, 
F.T.-.~Acosta$^{\rm 20}$, 
D.~Adamov\'{a}$^{\rm 93}$, 
A.~Adler$^{\rm 74}$, 
J.~Adolfsson$^{\rm 80}$, 
M.M.~Aggarwal$^{\rm 98}$, 
G.~Aglieri Rinella$^{\rm 34}$, 
M.~Agnello$^{\rm 31}$, 
N.~Agrawal$^{\rm 48}$, 
Z.~Ahammed$^{\rm 139}$, 
S.U.~Ahn$^{\rm 76}$
% ...

\bigskip

\bigskip 

\textbf{\Large Affiliation Notes}

\bigskip 

% ...
$^{\rm V}$ Also at: Institute of Theoretical Physics, University of Wroclaw, Poland\\

\bigskip

\bigskip 

\textbf{\Large Collaboration Institutes}

\bigskip 

% ...
$^{20}$ Department of Physics, University of California, Berkeley, California, United States\\
% ...
$^{31}$ Dipartimento DISAT del Politecnico and Sezione INFN, Turin, Italy\\
% ...
$^{34}$ European Organization for Nuclear Research (CERN), Geneva, Switzerland\\
% ...
$^{48}$ Indian Institute of Technology Bombay (IIT), Mumbai, India\\
% ...
$^{74}$ Johann-Wolfgang-Goethe Universit\"{a}t Frankfurt Institut f\"{u}r Informatik, Fachbereich Informatik und Mathematik, Frankfurt, Germany\\
% ...
$^{76}$ Korea Institute of Science and Technology Information, Daejeon, Republic of Korea\\
% ...
$^{80}$ Lund University Department of Physics, Division of Particle Physics, Lund, Sweden\\
% ...
$^{93}$ Nuclear Physics Institute of the Czech Academy of Sciences, \v{R}e\v{z} u Prahy, Czech Republic\\
% ...
$^{98}$ Physics Department, Panjab University, Chandigarh, India\\
% ...
$^{139}$ Variable Energy Cyclotron Centre, Homi Bhabha National Institute, Kolkata, India\\
% ...

\bigskip 

\end{flushleft} 

%% file: miniJetFF.bbl
\providecommand{\href}[2]{#2}\begingroup\raggedright\begin{thebibliography}{10}

\bibitem{dEnterria2014615}
D.~d'Enterria, K.~J. Eskola, I.~Helenius, and H.~Paukkunen, ``Confronting
  current {NLO} parton fragmentation functions with inclusive charged-particle
  spectra at hadron colliders,''
  \href{http://dx.doi.org/https://doi.org/10.1016/j.nuclphysb.2014.04.006}{{\em
  Nuclear Physics B} {\bfseries 883} (2014) 615 -- 628}.
  \url{http://www.sciencedirect.com/science/article/pii/S0550321314001151}.

\bibitem{ALICE_neutralMesons_pp7TeV}
{\bfseries ALICE} Collaboration, B.~Abelev {\em et~al.}, ``{Neutral pion and
  $\eta$ meson production in proton-proton collisions at $\sqrt{s}=0.9$ TeV and
  $\sqrt{s}=7$ TeV},''
  \href{http://dx.doi.org/10.1016/j.physletb.2012.09.015}{{\em Phys. Lett.}
  {\bfseries B717} (2012) 162--172},
\href{http://arxiv.org/abs/1205.5724}{{\ttfamily arXiv:1205.5724 [hep-ex]}}.
%%CITATION = ARXIV:1205.5724;%%.

\bibitem{Vogelsang_PhysRevD.48.3136}
L.~E. Gordon and W.~Vogelsang, ``Polarized and unpolarized prompt photon
  production beyond the leading order,''
  \href{http://dx.doi.org/10.1103/PhysRevD.48.3136}{{\em Phys. Rev. D}
  {\bfseries 48} (Oct, 1993) 3136--3159}.
  \url{https://link.aps.org/doi/10.1103/PhysRevD.48.3136}.

\bibitem{Minijets}
P.~Kotko, A.~M. Stasto, and M.~Strikman, ``{Exploring minijets beyond the
  leading power},''
\href{http://arxiv.org/abs/1608.00523}{{\ttfamily arXiv:1608.00523 [hep-ph]}}.
%%CITATION = ARXIV:1608.00523;%%.

\bibitem{A40_CTEQ5L}
{\bfseries CTEQ} Collaboration, H.~Lai {\em et~al.}, ``{Global QCD analysis of
  parton structure of the nucleon: CTEQ5 parton distributions},''
  \href{http://dx.doi.org/10.1007/s100529900196}{{\em Eur.Phys.J.} {\bfseries
  C12} (2000) 375--392},
\href{http://arxiv.org/abs/hep-ph/9903282}{{\ttfamily arXiv:hep-ph/9903282
  [hep-ph]}}.
%%CITATION = HEP-PH/9903282;%%.

\bibitem{powheg1}
{P. Nason}, ``{A New Method for Combining NLO QCD with Shower Monte Carlo
  Algorithms},'' {\em JHEP} {\bfseries 0411} (2004) 040.

\bibitem{powheg2}
{S. Frixione, P. Nason and C. Oleari}, ``{A general framework for implementing
  NLO calculations in shower Monte Carlo programs: the POWHEG BOX},'' {\em
  JHEP} {\bfseries 0711} (2007) 070.

\bibitem{powheg4}
S.~Alioli, K.~Hamilton, P.~Nason, C.~Oleari, and E.~Re, ``{Jet pair production
  in POWHEG},'' \href{http://dx.doi.org/10.1007/JHEP04(2011)081}{{\em JHEP}
  {\bfseries 04} (2011) 081},
\href{http://arxiv.org/abs/1012.3380}{{\ttfamily arXiv:1012.3380 [hep-ph]}}.
%%CITATION = ARXIV:1012.3380;%%.

\bibitem{Enterria_FF}
D.~d'Enterria, K.~J. Eskola, I.~Helenius, and H.~Paukkunen, ``{Confronting
  current NLO parton fragmentation functions with inclusive charged-particle
  spectra at hadron colliders},''
  \href{http://dx.doi.org/10.1016/j.nuclphysb.2014.04.006}{{\em Nucl. Phys.}
  {\bfseries B883} (2014) 615--628},
\href{http://arxiv.org/abs/1311.1415}{{\ttfamily arXiv:1311.1415 [hep-ph]}}.
%%CITATION = ARXIV:1311.1415;%%.

\bibitem{ATLAS_jets5fb}
{\bfseries ATLAS} Collaboration, G.~Aad {\em et~al.}, ``{Measurement of the
  inclusive jet cross-section in proton-proton collisions at $ \sqrt{s}=7 $ TeV
  using 4.5 fb$^{-1}$ of data with the ATLAS detector},''
  \href{http://dx.doi.org/10.1007/JHEP02(2015)153,
  10.1007/JHEP09(2015)141}{{\em JHEP} {\bfseries 02} (2015) 153},
  \href{http://arxiv.org/abs/1410.8857}{{\ttfamily arXiv:1410.8857 [hep-ex]}}.
[Erratum: JHEP09,141(2015)].
%%CITATION = ARXIV:1410.8857;%%.

\bibitem{A3_ATLASchJets}
{\bfseries ATLAS} Collaboration, {Aad, Georges and others}, ``{Properties of
  jets measured from tracks in proton-proton collisions at center-of-mass
  energy $\sqrt{s}=7$ TeV with the ATLAS detector},''
  \href{http://dx.doi.org/10.1103/PhysRevD.84.054001}{{\em Phys.Rev.}
  {\bfseries D84} (2011) 054001},
\href{http://arxiv.org/abs/1107.3311}{{\ttfamily arXiv:1107.3311 [hep-ex]}}.
%%CITATION = ARXIV:1107.3311;%%.

\bibitem{A5_CMSJets}
{\bfseries CMS} Collaboration, S.~Chatrchyan {\em et~al.}, ``{Measurement of
  the Inclusive Jet Cross Section in $pp$ Collisions at $\sqrt{s}=7$ TeV},''
  \href{http://dx.doi.org/10.1103/PhysRevLett.107.132001}{{\em Phys.Rev.Lett.}
  {\bfseries 107} (2011) 132001},
\href{http://arxiv.org/abs/1106.0208}{{\ttfamily arXiv:1106.0208 [hep-ex]}}.
%%CITATION = ARXIV:1106.0208;%%.

\bibitem{A5_CMSJets_1}
{\bfseries CMS} Collaboration, S.~Chatrchyan {\em et~al.}, ``{Measurements of
  differential jet cross sections in proton-proton collisions at $\sqrt{s}=7$
  TeV with the CMS detector},''
  \href{http://dx.doi.org/10.1103/PhysRevD.87.112002}{{\em Phys.Rev.}
  {\bfseries D87} no.~11, (2013) 112002},
\href{http://arxiv.org/abs/1212.6660}{{\ttfamily arXiv:1212.6660 [hep-ex]}}.
%%CITATION = ARXIV:1212.6660;%%.

\bibitem{A13_ATLAS_FF}
{\bfseries ATLAS} Collaboration, G.~Aad {\em et~al.}, ``{Measurement of the jet
  fragmentation function and transverse profile in proton-proton collisions at
  a center-of-mass energy of 7 TeV with the ATLAS detector},''
  \href{http://dx.doi.org/10.1140/epjc/s10052-011-1795-y}{{\em Eur.Phys.J.}
  {\bfseries C71} (2011) 1795},
\href{http://arxiv.org/abs/1109.5816}{{\ttfamily arXiv:1109.5816 [hep-ex]}}.
%%CITATION = ARXIV:1109.5816;%%.

\bibitem{ATLAS_PbPb_FF}
{\bfseries ATLAS} Collaboration, G.~Aad {\em et~al.}, ``{Measurement of
  inclusive jet charged-particle fragmentation functions in Pb+Pb collisions at
  $\sqrt{s_{NN}}$=2.76 TeV with the ATLAS detector},''
  \href{http://dx.doi.org/10.1016/j.physletb.2014.10.065}{{\em Phys. Lett.}
  {\bfseries B739} (2014) 320--342},
\href{http://arxiv.org/abs/1406.2979}{{\ttfamily arXiv:1406.2979 [hep-ex]}}.
%%CITATION = ARXIV:1406.2979;%%.

\bibitem{A14_CMS_PbPb_FF}
{\bfseries CMS} Collaboration, S.~Chatrchyan {\em et~al.}, ``{Measurement of
  jet fragmentation into charged particles in $pp$ and PbPb collisions at
  $\sqrt{s_{NN}}=2.76$ TeV},''
  \href{http://dx.doi.org/10.1007/JHEP10(2012)087}{{\em JHEP} {\bfseries 1210}
  (2012) 087},
\href{http://arxiv.org/abs/1205.5872}{{\ttfamily arXiv:1205.5872 [nucl-ex]}}.
%%CITATION = ARXIV:1205.5872;%%.

\bibitem{ALICE_chJets7TeV}
{\bfseries ALICE} Collaboration, B.~B. Abelev {\em et~al.}, ``{Charged jet
  cross sections and properties in proton-proton collisions at $\sqrt{s}=7$
  TeV},'' \href{http://dx.doi.org/10.1103/PhysRevD.91.112012}{{\em Phys. Rev.}
  {\bfseries D91} no.~11, (2015) 112012},
\href{http://arxiv.org/abs/1411.4969}{{\ttfamily arXiv:1411.4969 [nucl-ex]}}.
%%CITATION = ARXIV:1411.4969;%%.

\bibitem{A25_RefAntikt}
M.~Cacciari, G.~P. Salam, and G.~Soyez, ``{The Anti-k(t) jet clustering
  algorithm},'' \href{http://dx.doi.org/10.1088/1126-6708/2008/04/063}{{\em
  JHEP} {\bfseries 0804} (2008) 063},
\href{http://arxiv.org/abs/0802.1189}{{\ttfamily arXiv:0802.1189 [hep-ph]}}.
%%CITATION = ARXIV:0802.1189;%%.

\bibitem{Pythia_colorReconnections}
G.~Gustafson, ``{Multiple Interactions, Saturation, and Final States in pp
  Collisions and DIS},'' {\em Acta Phys. Polon.} {\bfseries B40} (2009)
  1981--1996,
\href{http://arxiv.org/abs/0905.2492}{{\ttfamily arXiv:0905.2492 [hep-ph]}}.
%%CITATION = ARXIV:0905.2492;%%.

\bibitem{Herwig_colorReconnections}
S.~Gieseke, C.~Rohr, and A.~Siodmok, ``{Colour reconnections in Herwig++},''
  \href{http://dx.doi.org/10.1140/epjc/s10052-012-2225-5}{{\em Eur. Phys. J.}
  {\bfseries C72} (2012) 2225},
\href{http://arxiv.org/abs/1206.0041}{{\ttfamily arXiv:1206.0041 [hep-ph]}}.
%%CITATION = ARXIV:1206.0041;%%.

\bibitem{A20_AliceExpt}
{\bfseries ALICE} Collaboration, K.~Aamodt {\em et~al.}, ``{The ALICE
  experiment at the CERN LHC},''
\href{http://dx.doi.org/10.1088/1748-0221/3/08/S08002}{{\em JINST} {\bfseries
  3} (2008) S08002}.
%%CITATION = JINST,3,S08002;%%.

\bibitem{A21_RefTPC}
J.~Alme, Y.~Andres, H.~Appelshauser, S.~Bablok, N.~Bialas, {\em et~al.}, ``{The
  ALICE TPC, a large 3-dimensional tracking device with fast readout for
  ultra-high multiplicity events},''
  \href{http://dx.doi.org/10.1016/j.nima.2010.04.042}{{\em Nucl.Instrum.Meth.}
  {\bfseries A622} (2010) 316--367},
\href{http://arxiv.org/abs/1001.1950}{{\ttfamily arXiv:1001.1950
  [physics.ins-det]}}.
%%CITATION = ARXIV:1001.1950;%%.

\bibitem{A22_RefITS}
{\bfseries ALICE} Collaboration, K.~Aamodt {\em et~al.}, ``{Alignment of the
  ALICE Inner Tracking System with cosmic-ray tracks},''
  \href{http://dx.doi.org/10.1088/1748-0221/5/03/P03003}{{\em JINST} {\bfseries
  5} (2010) P03003},
\href{http://arxiv.org/abs/1001.0502}{{\ttfamily arXiv:1001.0502
  [physics.ins-det]}}.
%%CITATION = ARXIV:1001.0502;%%.

\bibitem{A23_RefVZERO}
{\bfseries ALICE} Collaboration, P.~Cortese {\em et~al.}, ``{ALICE technical
  design report on forward detectors: FMD, T0 and V0},''
{\em CERN-LHCC-2004-025, https://cds.cern.ch/record/781854} (2004) .
%%CITATION = CERN-LHCC-2004-025 ETC.;%%.

\bibitem{A24_AliceSigmaPaper}
{\bfseries ALICE} Collaboration, B.~Abelev {\em et~al.}, ``{Measurement of
  inelastic, single- and double-diffraction cross sections in proton--proton
  collisions at the LHC with ALICE},''
  \href{http://dx.doi.org/10.1140/epjc/s10052-013-2456-0}{{\em Eur.Phys.J.}
  {\bfseries C73} (2013) 2456},
\href{http://arxiv.org/abs/1208.4968}{{\ttfamily arXiv:1208.4968 [hep-ex]}}.
%%CITATION = ARXIV:1208.4968;%%.

\bibitem{A15_FullJetPaper}
{\bfseries ALICE} Collaboration, B.~Abelev {\em et~al.}, ``{Measurement of the
  inclusive differential jet cross section in $pp$ collisions at $\sqrt{s} =
  2.76$ TeV},'' \href{http://dx.doi.org/10.1016/j.physletb.2013.04.026}{{\em
  Phys.Lett.} {\bfseries B722} (2013) 262--272},
\href{http://arxiv.org/abs/1301.3475}{{\ttfamily arXiv:1301.3475 [nucl-ex]}}.
%%CITATION = ARXIV:1301.3475;%%.

\bibitem{A27_RefFastjet}
M.~Cacciari and G.~P. Salam, ``{Dispelling the $N^{3}$ myth for the $k_t$
  jet-finder},'' \href{http://dx.doi.org/10.1016/j.physletb.2006.08.037}{{\em
  Phys.Lett.} {\bfseries B641} (2006) 57--61},
\href{http://arxiv.org/abs/hep-ph/0512210}{{\ttfamily arXiv:hep-ph/0512210
  [hep-ph]}}.
%%CITATION = HEP-PH/0512210;%%.

\bibitem{Pythia}
{Sjostrand, Torbjorn and Mrenna, Stephen and Skands, Peter Z.}, ``{PYTHIA 6.4
  Physics and Manual},''
  \href{http://dx.doi.org/10.1088/1126-6708/2006/05/026}{{\em JHEP} {\bfseries
  0605} (2006) 026},
\href{http://arxiv.org/abs/hep-ph/0603175}{{\ttfamily arXiv:hep-ph/0603175
  [hep-ph]}}.
%%CITATION = HEP-PH/0603175;%%.

\bibitem{A31_PerugiaTunes}
P.~Z. Skands, ``{Tuning Monte Carlo Generators: The Perugia Tunes},''
  \href{http://dx.doi.org/10.1103/PhysRevD.82.074018}{{\em Phys.Rev.}
  {\bfseries D82} (2010) 074018},
\href{http://arxiv.org/abs/1005.3457}{{\ttfamily arXiv:1005.3457 [hep-ph]}}.
%%CITATION = ARXIV:1005.3457;%%.

\bibitem{A32_RefGeant3}
R.~Brun, F.~Carminati, and S.~Giani, ``{GEANT Detector Description and
  Simulation Tool},''
{\em CERN-W5013, CERN-W-5013, https://cds.cern.ch/record/1082634} .
%%CITATION = CERN-W5013 ETC.;%%.

\bibitem{Herwig}
G.~Marchesini, B.~Webber, G.~Abbiendi, I.~Knowles, M.~Seymour, {\em et~al.},
  ``{HERWIG: A Monte Carlo event generator for simulating hadron emission
  reactions with interfering gluons. Version 5.1 - April 1991},''
\href{http://dx.doi.org/10.1016/0010-4655(92)90055-4}{{\em Comput.Phys.Commun.}
  {\bfseries 67} (1992) 465--508}.
%%CITATION = CPHCB,67,465;%%.

\bibitem{A36_Herwig_1}
G.~Corcella, I.~Knowles, G.~Marchesini, S.~Moretti, K.~Odagiri, {\em et~al.},
  ``{HERWIG 6: An Event generator for hadron emission reactions with
  interfering gluons (including supersymmetric processes)},''
  \href{http://dx.doi.org/10.1088/1126-6708/2001/01/010}{{\em JHEP} {\bfseries
  0101} (2001) 010},
\href{http://arxiv.org/abs/hep-ph/0011363}{{\ttfamily arXiv:hep-ph/0011363
  [hep-ph]}}.
%%CITATION = HEP-PH/0011363;%%.

\bibitem{Pythia8}
T.~Sjostrand, S.~Mrenna, and P.~Z. Skands, ``{A Brief Introduction to PYTHIA
  8.1},'' \href{http://dx.doi.org/10.1016/j.cpc.2008.01.036}{{\em Comput. Phys.
  Commun.} {\bfseries 178} (2008) 852--867},
\href{http://arxiv.org/abs/0710.3820}{{\ttfamily arXiv:0710.3820 [hep-ph]}}.
%%CITATION = ARXIV:0710.3820;%%.

\bibitem{A37_pt_ordered_showers}
T.~Sjostrand and P.~Z. Skands, ``{Transverse-momentum-ordered showers and
  interleaved multiple interactions},''
  \href{http://dx.doi.org/10.1140/epjc/s2004-02084-y}{{\em Eur.Phys.J.}
  {\bfseries C39} (2005) 129--154},
\href{http://arxiv.org/abs/hep-ph/0408302}{{\ttfamily arXiv:hep-ph/0408302
  [hep-ph]}}.
%%CITATION = HEP-PH/0408302;%%.

\bibitem{A38_Lund_model}
B.~Andersson, G.~Gustafson, and B.~Soderberg, ``{A General Model for Jet
  Fragmentation},''
\href{http://dx.doi.org/10.1007/BF01407824}{{\em Z.Phys.} {\bfseries C20}
  (1983) 317}.
%%CITATION = ZEPYA,C20,317;%%.

\bibitem{Pythia8Monash}
P.~Skands, S.~Carrazza, and J.~Rojo, ``{Tuning PYTHIA 8.1: the Monash 2013
  Tune},'' \href{http://dx.doi.org/10.1140/epjc/s10052-014-3024-y}{{\em Eur.
  Phys. J.} {\bfseries C74} no.~8, (2014) 3024},
\href{http://arxiv.org/abs/1404.5630}{{\ttfamily arXiv:1404.5630 [hep-ph]}}.
%%CITATION = ARXIV:1404.5630;%%.

\bibitem{PDF_NNPDF}
{\bfseries NNPDF} Collaboration, R.~D. Ball, V.~Bertone, S.~Carrazza,
  L.~Del~Debbio, S.~Forte, A.~Guffanti, N.~P. Hartland, and J.~Rojo, ``{Parton
  distributions with QED corrections},''
  \href{http://dx.doi.org/10.1016/j.nuclphysb.2013.10.010}{{\em Nucl. Phys.}
  {\bfseries B877} (2013) 290--320},
\href{http://arxiv.org/abs/1308.0598}{{\ttfamily arXiv:1308.0598 [hep-ph]}}.
%%CITATION = ARXIV:1308.0598;%%.

\bibitem{PDF_CTEQ6}
J.~Pumplin, D.~R. Stump, J.~Huston, H.~L. Lai, P.~M. Nadolsky, and W.~K. Tung,
  ``{New generation of parton distributions with uncertainties from global QCD
  analysis},'' \href{http://dx.doi.org/10.1088/1126-6708/2002/07/012}{{\em
  JHEP} {\bfseries 07} (2002) 012},
\href{http://arxiv.org/abs/hep-ph/0201195}{{\ttfamily arXiv:hep-ph/0201195
  [hep-ph]}}.
%%CITATION = HEP-PH/0201195;%%.

\bibitem{LHAPDF5}
M.~R. Whalley, D.~Bourilkov, and R.~C. Group, ``{The Les Houches accord PDFs
  (LHAPDF) and LHAGLUE},'' in {\em {HERA and the LHC: A Workshop on the
  implications of HERA for LHC physics. Proceedings, Part B}}, pp.~575--581.
\newblock 2005.
\newblock
\href{http://arxiv.org/abs/hep-ph/0508110}{{\ttfamily arXiv:hep-ph/0508110
  [hep-ph]}}.
\newblock
%%CITATION = HEP-PH/0508110;%%.

\bibitem{A43_unfold-bayes}
G.~D'Agostini, ``{A Multidimensional unfolding method based on Bayes'
  theorem},''
\href{http://dx.doi.org/10.1016/0168-9002(95)00274-X}{{\em Nucl.Instrum.Meth.}
  {\bfseries A362} (1995) 487--498}.
%%CITATION = NUIMA,A362,487;%%.

\bibitem{A45_RooUnfoldHtml}
T.~Adye, ``{Unfolding algorithms and tests using RooUnfold},''
  \href{http://arxiv.org/abs/1105.1160}{{\ttfamily arXiv:1105.1160
  [physics.data-an]}}.
See also http://hepunx.rl.ac.uk/~adye/software/unfold/RooUnfold.html.
%%CITATION = ARXIV:1105.1160;%%.

\bibitem{A44_unfold-svd}
A.~Hocker and V.~Kartvelishvili, ``{SVD approach to data unfolding},''
  \href{http://dx.doi.org/10.1016/0168-9002(95)01478-0}{{\em
  Nucl.Instrum.Meth.} {\bfseries A372} (1996) 469--481},
\href{http://arxiv.org/abs/hep-ph/9509307}{{\ttfamily arXiv:hep-ph/9509307
  [hep-ph]}}.
%%CITATION = HEP-PH/9509307;%%.

\bibitem{A46_CMSStrangeness}
{\bfseries CMS} Collaboration, V.~Khachatryan {\em et~al.}, ``{Strange Particle
  Production in $pp$ Collisions at $\sqrt{s}=0.9$ and 7 TeV},''
  \href{http://dx.doi.org/10.1007/JHEP05(2011)064}{{\em JHEP} {\bfseries 1105}
  (2011) 064},
\href{http://arxiv.org/abs/1102.4282}{{\ttfamily arXiv:1102.4282 [hep-ex]}}.
%%CITATION = ARXIV:1102.4282;%%.

\bibitem{Karneyeu:2013aha}
A.~Karneyeu, L.~Mijovic, S.~Prestel, and P.~Z. Skands, ``{MCPLOTS: a particle
  physics resource based on volunteer computing},''
  \href{http://dx.doi.org/10.1140/epjc/s10052-014-2714-9}{{\em Eur. Phys. J.}
  {\bfseries C74} (2014) 2714},
\href{http://arxiv.org/abs/1306.3436}{{\ttfamily arXiv:1306.3436 [hep-ph]}}.
%%CITATION = ARXIV:1306.3436;%%.

\bibitem{A47_mcplots}
P.~Skands {\em et~al.} {\em https://mcplots.cern.ch} .

\bibitem{A52_R_AA_momentumResolution}
{\bfseries ALICE} Collaboration, B.~Abelev {\em et~al.}, ``{Centrality
  Dependence of Charged Particle Production at Large Transverse Momentum in
  Pb--Pb Collisions at $\sqrt{s_{\rm{NN}}} = 2.76$ TeV},''
  \href{http://dx.doi.org/10.1016/j.physletb.2013.01.051}{{\em Phys.Lett.}
  {\bfseries B720} (2013) 52--62},
\href{http://arxiv.org/abs/1208.2711}{{\ttfamily arXiv:1208.2711 [hep-ex]}}.
%%CITATION = ARXIV:1208.2711;%%.

\bibitem{UA1_minijets}
{\bfseries UA1} Collaboration, C.~Albajar {\em et~al.}, ``{Production of Low
  Transverse Energy Clusters in anti-p p Collisions at s**(1/2) = 0.2-TeV to
  0.9-TeV and their Interpretation in Terms of QCD Jets},''
\href{http://dx.doi.org/10.1016/0550-3213(88)90450-6}{{\em Nucl. Phys.}
  {\bfseries B309} (1988) 405--425}.
%%CITATION = NUPHA,B309,405;%%.

\bibitem{A51_ALICE_UE}
{\bfseries ALICE} Collaboration, B.~Abelev {\em et~al.}, ``{Underlying Event
  measurements in $pp$ collisions at $\sqrt{s}=0.9$ and 7 TeV with the ALICE
  experiment at the LHC},''
  \href{http://dx.doi.org/10.1007/JHEP07(2012)116}{{\em JHEP} {\bfseries 1207}
  (2012) 116},
\href{http://arxiv.org/abs/1112.2082}{{\ttfamily arXiv:1112.2082 [hep-ex]}}.
%%CITATION = ARXIV:1112.2082;%%.

\bibitem{CDF_FF}
{\bfseries CDF} Collaboration, {Acosta, D. and others}, ``{Momentum
  distribution of charged particles in jets in dijet events in $p \bar{p}$
  collisions at $\sqrt{s} = 1.8$ TeV and comparisons to perturbative QCD
  predictions},''
\href{http://dx.doi.org/10.1103/PhysRevD.68.012003}{{\em Phys.Rev.} {\bfseries
  D68} (2003) 012003}.
%%CITATION = PHRVA,D68,012003;%%.

\bibitem{A29_QCD_coherence}
B.~Ermolaev and V.~S. Fadin, ``{Log - Log Asymptotic Form of Exclusive
  Cross-Sections in Quantum Chromodynamics},''
{\em JETP Lett.} {\bfseries 33} (1981) 269--272.
%%CITATION = JTPLA,33,269;%%.

\bibitem{A29_QCD_coherence_1}
A.~H. Mueller, ``{On the Multiplicity of Hadrons in QCD Jets},''
\href{http://dx.doi.org/10.1016/0370-2693(81)90581-5}{{\em Phys.Lett.}
  {\bfseries B104} (1981) 161--164}.
%%CITATION = PHLTA,B104,161;%%.

\bibitem{Ortiz_colorReconnections}
A.~Ortiz~Velasquez, P.~Christiansen, E.~Cuautle~Flores, I.~Maldonado~Cervantes,
  and G.~Paić, ``{Color Reconnection and Flowlike Patterns in $pp$
  Collisions},'' \href{http://dx.doi.org/10.1103/PhysRevLett.111.042001}{{\em
  Phys. Rev. Lett.} {\bfseries 111} no.~4, (2013) 042001},
\href{http://arxiv.org/abs/1303.6326}{{\ttfamily arXiv:1303.6326 [hep-ph]}}.
%%CITATION = ARXIV:1303.6326;%%.

\end{thebibliography}\endgroup
